\documentclass[
aps,%
12pt,%
final,%
notitlepage,%
oneside,%
onecolumn,%
nobibnotes,%
nofootinbib,%
superscriptaddress,%
noshowpacs]%
{revtex4}%
\usepackage{graphics}
\usepackage[T2A]{fontenc}
\usepackage[english]{babel}
\usepackage{amssymb,amsmath}

\begin{document}

\title{Production  of double charmed baryons with the excited heavy diquark at LHC}
\author{\firstname{A.~V.}~\surname{Berezhnoy}}
\email{Alexander.Berezhnoy@cern.ch}
\affiliation{SINP of Moscow State University, Russia}

\author{\firstname{I.~N.}~\surname{Belov}}
\email{iljattbelov@gmail.com}
\affiliation{Physical Department of Moscow State University, Russia}

\author{\firstname{A.~K.}~\surname{Likhoded}}
\email{Anatolii.Likhoded@ihep.ru}
\affiliation{"Institute for High Energy Physics" NRC "Kurchatov Institute", 142281, Protvino, Russia }

\begin{abstract}
The yield of doubly charmed baryons with excited   heavy diquark in $S$ wave and $P$ wave states  has been estimated at LHC energies. The observation possibility of such baryons is discussed. 
\end{abstract} 

\maketitle

\section{introduction}

The problems of production and decays  of the doubly heavy baryons was of interest to researchers for many years.   Such systems consist of   two charm quarks and one light quark, and,   therefore, it is quite natural to divide calculating the characteristics of doubly heavy baryon in two stages: the calculation of the properties of the heavy diquark and the subsequent calculation of the properties of  quark-diquark system. This essentially simplifies a theoretical research of  doubly heavy baryons, and allows to obtain  the detailed prediction of their properties (see, for example\cite{Ebert:1996ec,Kiselev:2001fw,Faustov:2018vgl}). It is necessary to note that there are attempts to study doubly heavy baryons spectroscopy by direct solving  of the quantum three body problem~(see, for example, \cite{Kerbikov:1987vx, Fleck:1989mb, Roncaglia:1995az, SilvestreBrac:1996bg,Albertus:2006wb, Albertus:2006ya,Roberts:2007ni,Yoshida:2015tia}). 
 Thus, the spectroscopy of doubly heavy baryons can be investigated within a three body potential model, as well as within a quark-diquark approach. But studying the production of doubly heavy baryons does not provide such a choice. The only more or less consistent model of doubly heavy baryon production known so far is based on the assumption that the originally produced doubly heavy diquark transforms to the doubly heavy baryon.

\begin{figure}
 \centering
\resizebox*{0.6\textwidth}{!}{\includegraphics{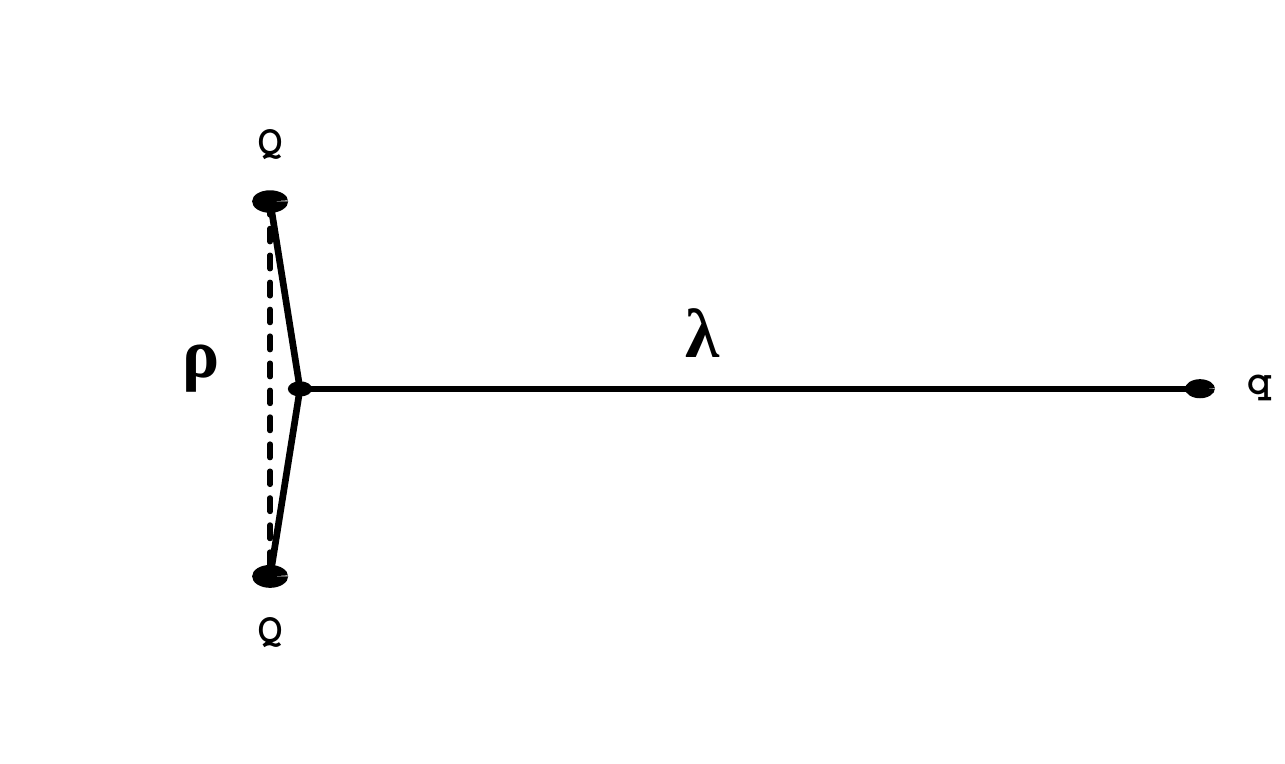}}
\caption{ Schematic representation of $\rho$ and $\lambda$ excited states of $\Xi_{cc}$ baryon. $\rho$ states are states with the excited doubly heavy diquark, $\lambda$ are states with the excited light quark.}
\label{fig:Xicc_excitations}
\end{figure}

For many years, these particles could not be observed experimentally. But finally the first doubly heavy baryon $\Xi_{cc}^{++}$ was observed by the LHCb Collaboration in the decay mode $\Lambda_c^+ K^- \pi^+ \pi^+$ \cite{Aaij:2017ueg}. The observation  has been already  confirmed in the mode $\Xi_c^+ \pi^+$ \cite{Aaij:2018gfl}. The lifetime of this new state also has been measured \cite{Aaij:2018wzf}. This circumstance greatly revived the research activities in this direction. In this article we discuss the possibilities of further research of doubly charm baryon states, namely we estimate within the quark-diquark appoarch the yield of the doubly charmed baryons with excited  heavy diquark (so-called $\rho$-exitations, see Fig.~\ref{fig:Xicc_excitations}).

\section{production}    

To produce  a baryon, it is natural to use a two-step procedure. In the first step of the calculations a double heavy diquark in the anti-triplet color state can be produced perturbatively in the hard  interaction: in the second step a double heavy diquark should be transformed to the baryon within the soft hadronization process (see \cite{Berezhnoy:1995fy, Baranov:1995rc, Baranov:1997sg, Berezhnoy:1998aa, Chang:2006eu} for details).

The production amplitude can be written as follows:\begin{equation}
A^{SJj_z}=\int T^{Ss_z}_{c\bar c c \bar c}(p_i,k(\vec q))\cdot
\left (\Psi^{Ll_z}_{ [c c]_{\bar 3_c}}(\vec q) \right )^* \cdot
C^{Jj_z}_{s_zl_z} \frac{d^3 \vec q}{{(2\pi)}^3},
\end{equation}
where $T^{Ss_z}_{c\bar c c \bar c}$ is an amplitude of the hard production of two heavy quark pairs; $\Psi^{Ll_z}_{[cc]_{\bar 3_c}}$ is a wave function of the diquark color antitriplet; $J$ and $j_z$ are the total angular momentum and its projection on $z$-axis in the $[cc]_{\bar 3_c}$ diquark rest frame; $L$ and $l_z$ are the orbital angular momentum of $[cc]_{\bar 3_c}$ diquark and its projection on $z$-axis; $S$ and $s_z$ are $cc$-diquark spin and its projection; $C^{Jj_z}_{s_zl_z}$ are Clebsh-Gordon coefficients; $p_i$ are four momenta of  diquark,  and $\bar c$ quarks; $\vec q$  is three momentum of $c$-quark in the  $[cc]_{\bar 3_c}$ diquark rest frame (in this frame $(0,\vec q ) = k(\vec q)$).

Under assumption of small dependence of $T^{Ss_z}_{c \bar c c \bar c}$ on $k(\vec q)$ amplitude can be expanded into a series of $\vec q$ powers:
\begin{equation}
A \sim 
\int d^3q\,\Psi^*({\vec q})\left\{ \bigl.T(p_i,{\vec q}) \bigr|_{\vec q=0}+
\bigl.{\vec q}\frac{\partial}{\partial {\vec q}} T(p_i,\vec q) \bigr|_{\vec q =0}  +  \dotsb \right\},
\end{equation}
where the first term provides us the $S$-wave matrix element, the second term --- $P$-wave. 

The amplitude $T^{Ss_z}_{c\bar c c \bar c}$ is calculated numerically within LO of perturbative QCD. The two quarks are combined into the color antitriplet diquark with the given spin value (see \cite{Berezhnoy:1998aa} for details). The derivatives on $\vec q$   are also calculated numerically, as it was done calculating matrix elements of $P$-wave $B_c$  production~\cite{Berezhnoy:1996ks}. 

Since the spectroscopy of a diquark with two identical quarks puts a restriction on the spin $S$ of a diquark, the formulae are simplified. For $S$-wave state the diquark spin $S=1$ and  $j_z=s_z$:  
\begin{equation}
A^{s_z} = \frac{1}{\sqrt{4\pi}}R_S(0) \cdot \bigl. T_{c\bar{c}c\bar{c}}^{s_z}(p_i) \bigr|_{\vec q=0}. 
\end{equation}
For $P$-wave state the diquark spin  $S=0$ and $j_z=l_z$: 
\begin{equation}
A^{l_z} = i\sqrt{\frac{3}{4\pi}} R_P'(0) \cdot \bigl.\{{\cal L}^{l_z} T_{c\bar{c}c\bar{c}}\left(p_i, \vec{q}\right )\}\bigr|_{\vec q=0}, 
\end{equation}
where $R_S(0)$ and $R_P'(0)$ are values of radial wave function at origin; ${\cal L}^{l_z}$ is a differential operator of the following form:

\begin{equation}
{\cal L}^{l_z} = \begin{cases}
{\cal L}^{-1}=\frac{1}{\sqrt{2}}\left (\frac{\partial}{\partial q_x}
+i\frac{\partial}{\partial q_y} \right ) \\
{\cal L}^0=\frac{\partial}{\partial q_z} \\
{\cal L}^{+1}=-\frac{1}{\sqrt{2}}\left (\frac{\partial}{\partial q_x}
-i\frac{\partial}{\partial q_y} \right )\\
\end{cases}
\end{equation}

Obviously, a color antitriplet of $cc$ system  should be somehow transformed to the  $ccq$ baryon.  The transverse momentum of light quark $q$  with mass $m_q$ is about
$\frac{m_q}{M}p_T$, where $M$ is a mass of $(ccq)$-baryon, and $p_T$ is its transverse momentum.  For LHCb kinematical conditions such a quark always exists in the quark sea. In our estimations we assume, that a doubly heavy diquark is hadronized by joining  with one of the light quarks $u$, $d$ and $s$ in the same proportion, as a $b$ quark:  $1:1:0.26$~\cite{Aaij:2011jp}. We also assume that  it is hadronized with unite probability. The latter assumption is  pretty much a guess,  because diquark has a color charge and therefore strongly interacts with its  environment,  that could lead to the diquark dissociation.


For our estimations we use wave function values from \cite{Ebert:2002ig} (see Table~\ref{parameters}). To obtain the proton-proton cross sections we use  PDFs  and $\alpha_s$ from CT14 PDF set \cite{Dulat:2015mca}. The calculations are performed for LHCb detector's kinematics $ 2<\eta<4.5,\ p_T<10$ GeV at center-of-mass energy $\sqrt{s}=13$ TeV, where $\eta$ is a baryon pseudorapidity and $p_T$ is its transverse momentum. The scale variation on baryon transverse energy $E_T= \sqrt{M^2+p_T^2}$ from $E_T/2$ to $2E_T$ contributes to uncertainty for cross sections. Final results are given in Table~\ref{results}, where by relative yield we mean the ratio of cross sections. 
The relative yield of baryons with doubly charmed diquark in excited $2S$ and $3S$ states is about 50\%.   $P$~wave  states of the diquark give only $3\div5$ \% of the  total yield.

\begin{figure}[ht]
 \centering
\resizebox*{0.7\textwidth}{!}{\includegraphics{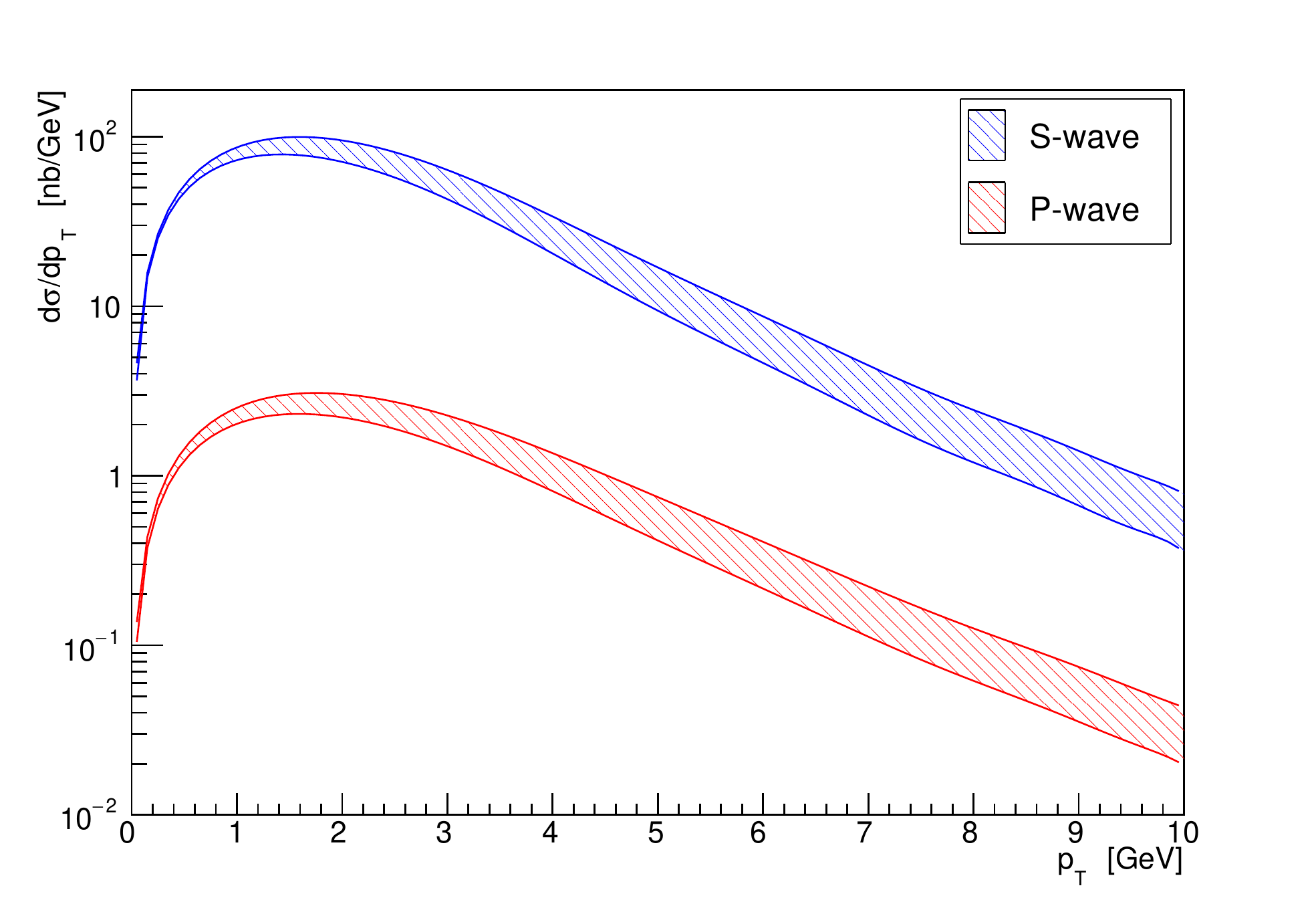}}
\caption{ $d\sigma/d p_T$ dependence on $p_T$ for different scales.}
\label{fig:SPLogScale}
\end{figure}

\begin{figure}[ht]
 \centering
\resizebox*{0.7\textwidth}{!}{\includegraphics{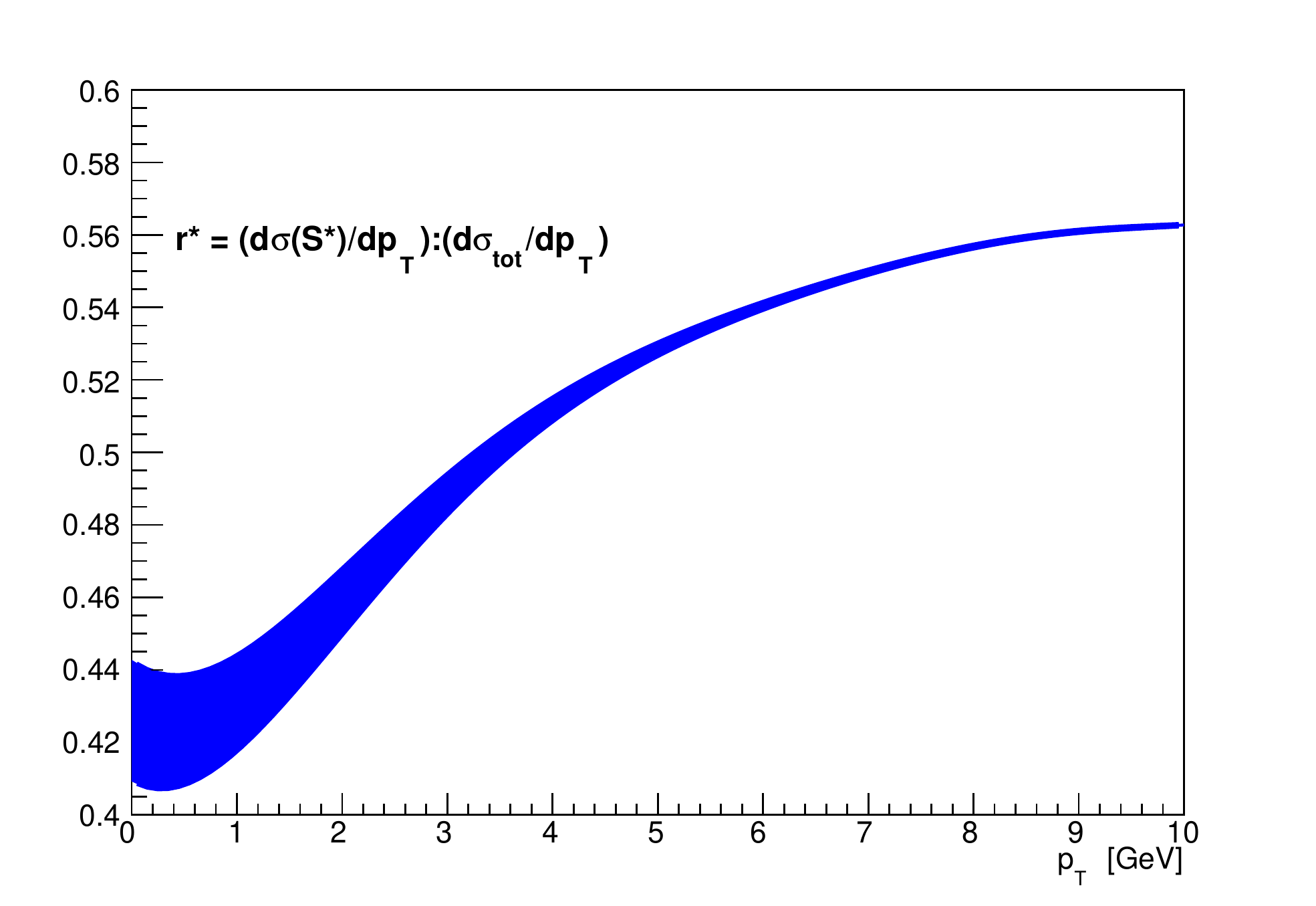}}
\caption{ $r^*$ dependence on $p_T$ for different scales.}
\label{fig:Pt_ratio_S}
\end{figure}

\begin{figure}[ht]
 \centering
\resizebox*{0.7\textwidth}{!}{\includegraphics{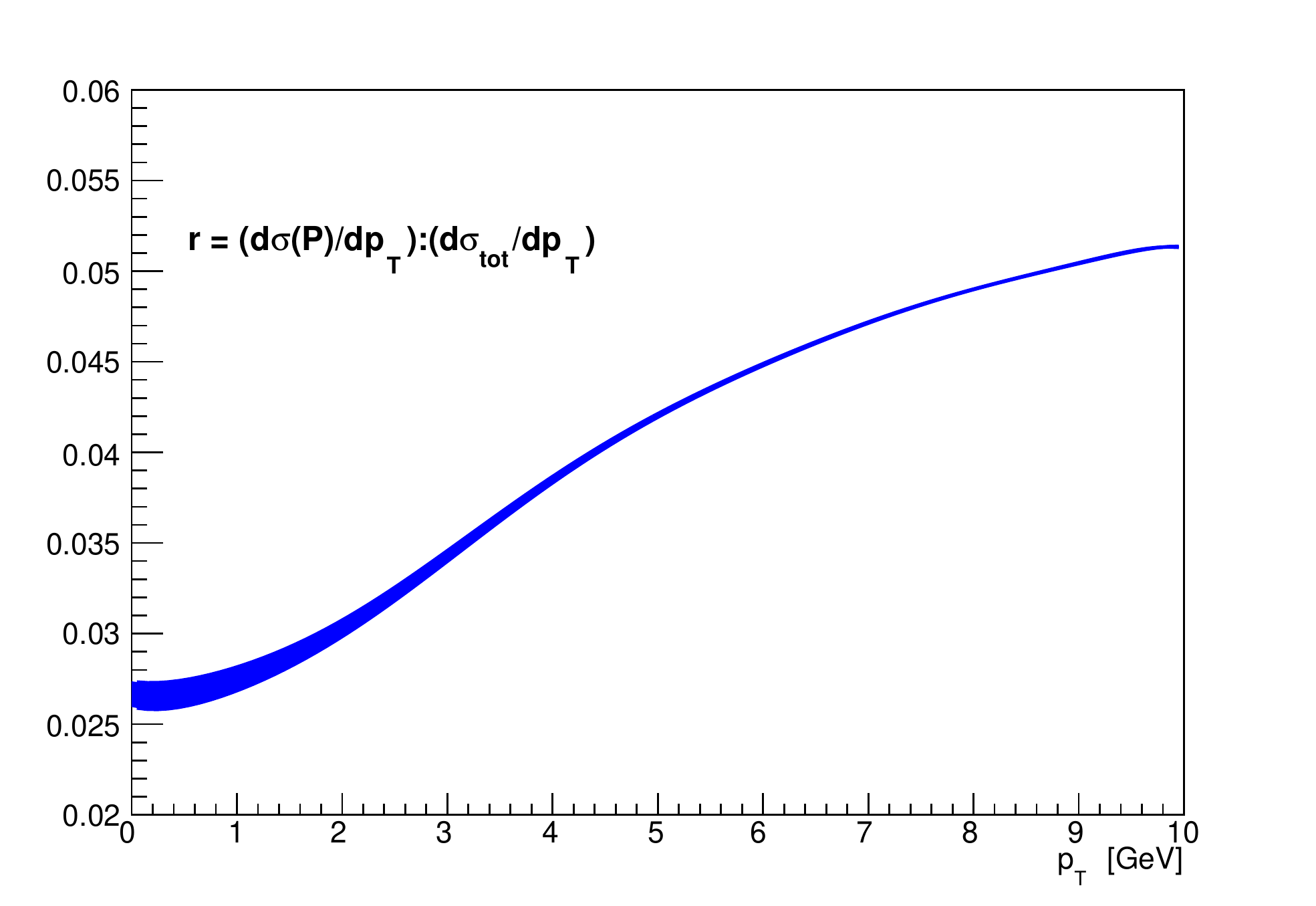}}
\caption{ $r$ dependence on $p_T$ for different scales.}
\label{fig:Pt_ratio_P}
\end{figure}

The $p_T$ distributions at $\sqrt{s}=13$ TeV are presented in Fig.~\ref{fig:SPLogScale}.
 It can be seen in Figs. \ref{fig:Pt_ratio_S} and \ref{fig:Pt_ratio_P}, that at large transverse momenta  the  uncertainties for the relative values arising from the scale choice are practically cancel out. At small transverse momenta these uncertainties become more sizable achieving $\sim 10$\%.

Our estimations show that the relative contribution of the excited states slightly increases with transverse momenta. However it doesn't mean, that excitations should be sought  at large transverse momenta, because an absolute yield is greater at small transverse momenta.

The $p_T$ distributions and yield ratios at $\sqrt{s}=8$ TeV are very similar to the corresponding distributions at $\sqrt{s}=13$ TeV and are therefore not presented here.

\begin{table}[ht]
\caption{Wave function values and masses for the doubly charmed diquark~\cite{Ebert:2002ig}.}
\label{parameters}
\begin{tabular}{|c|c|c|}\hline
state & wave function  & diquark mass \\
\hline
 & $|R(0)|$, GeV$^{3/2}$ & $m$, GeV  \\
$1S$  & 0.566 &  3.20\\
$2S$  & 0.540 &  3.50\\
$3S$  & 0.542 &  3.70\\\hline 
 & $|R'(0)|$, GeV$^{5/2}$ & $m$, GeV \\
$1P$  & 0.149 &  3.40\\
$2P$  &  0.198 &  3.70\\
\hline
\end{tabular}
\end{table}

\begin{table}[ht]
\caption{Cross sections and relative yields for $cc$-diquark states.}
\label{results}
\begin{tabular}{|c|c|c|}\hline
state & relative yield  & cross section \\
\hline
 & $r^*$, \% & $\sigma$, nb  \\
$1S$  & $49\div52$ & $120\div170$ \\
$2S$  & $26\div27$ & $60\div90$ \\
$3S$  & $18\div20$ & $40\div70$\\
\hline 
 & $r$, \% & $\sigma$, nb  \\
$1P$  & 2 & $4\div6$\\
$2P$  &  $1\div2$ & $4\div5$\\
\hline
\end{tabular}
\end{table}

\newpage

\newpage
\section{Transitions of doubly charmed baryons to the ground state}

Here we briefly review the current state of theoretical research on decays of doubly charmed excited baryons.

The excited states of doubly charmed baryons, lying below the $\Lambda_c D$ threshold, fall into the ground state. It is anticipated that, if kinematically possible, the hadronic mode as a rule dominates (see \cite{Dai:2000hza,Lu:2017meb,Xiao:2017udy} for electromagnetic transitions and \cite{Ma:2017nik,Ma:2015lba,Xiao:2017dly,Xiao:2017udy,Mehen:2017nrh,Hu:2005gf} for the hadronic ones).


The quark-diquark model of a doubly heavy baryon allows one to examine separately the excitations of a light quark and a heavy diquark. Therefore, transitions between the different states of doubly heavy baryon can be categorized into transitions caused by a change of the light quark state in the baryon and transitions caused by a change of the diquark state.
All research groups predict that $\lambda$-excitations of doubly charmed baryons should be rather broad (40-300 MeV, see~\cite{Xiao:2017dly,Xiao:2017udy,Ma:2017nik,Ma:2015lba}).


As for $\rho$ excitations, the production of which we study, there is a disagreement in predictions.
As predicted in~\cite{Mehen:2017nrh},  the decay widths of doubly charmed baryons with  first radial  excitation of the diquark  are comparable in magnitude to the values in case of transitions from excitations of light quark degrees of freedom: $\Gamma\left(\Xi_{cc}(2S1s(1/2))\right) \sim 50\ \text{MeV}$\footnote{Hereinafter we use a number with an uppercase letter for the heavy diquark orbital state, a number with a lowercase letter for the light quark orbital state, and a number in parentheses for the total angular momentum of the baryon.} and $\Gamma\left(\Xi_{cc}(2S1s(3/2))\right) \sim 400\ \text{MeV}$,
where the decays go into either levels of doublet $1S1s$.
However, the results of  \cite{Mehen:2017nrh} contradict \cite{Eakins:2012fq}, where the values less than 0.5 MeV are predicted for the transition widths $\Xi_{cc}(2S)\to \Xi_{cc}(1S)\pi$.

It was discussed in~\cite{Gershtein:1998un},  that doubly charmed baryons with the $P$ wave heavy diquark state should be metastable, because their decay widths are  suppressed as $\Lambda_{QCD}^2/m_c^2$ since such transitions are accompanied by a simultaneous change of the spin and orbital angular momentum of the diquark.
In \cite{Hu:2005gf}  the decay widths of $\Xi_{cc}(1P)$ states have been estimated as follows:
\begin{equation}
\begin{cases}
&\Gamma\left[\Xi_{cc}(1P1s(3/2)) \to \Xi_{cc}(1S1s(3/2)) \pi\right] = \lambda_{3/2}^{2} 112\ \text{MeV},\\  
&\Gamma\left[\Xi_{cc}(1P1s(1/2))\to \Xi_{cc}(1S1s(1/2)) \pi \right] = \lambda_{1/2}^{2} 111\ \text{MeV},
\end{cases}
\end{equation}
where $\lambda_{3/2},\lambda_{1/2} \sim \Lambda_{QCD}/m_c$. Thus, for reasonable values of   $\lambda_{1/2}$ and $\lambda_{3/2}$  these states will indeed be  metastable, as supposed in~\cite{Gershtein:1998un}.

The decay $\Xi_{cc}(1P1s(1/2))\to \Xi_{cc}(1S1s(1/2)) \pi$ can be fully reconstracted. The decay $\Xi_{cc}(1P1s(3/2)) \to \Xi_{cc}(1S1s(3/2)) \pi \to [ \Xi_{cc}(1S1s(1/2)) \gamma] \pi$ is likely to be reconstructed with the loss of the photon, because such relatively soft photon has a small registration efficiency.   However the peak corresponding to $\Xi_{cc}(1P1s(3/2))$ could be seen in  $\Xi_{cc}\pi$ mass distribution. This peak will be shifted by the value of   mass splitting in $1S1s$-doublet and   will get an additional width $\sim 10$ MeV. Thus, the loss of the  photon will not wash out peaks in  $\Xi_{cc}\pi$ mass spectrum.~\footnote{The additional width approximately equals $
 2\Delta M^S\sqrt{\left(\Delta M^{PS}/M\right)^2 - \left(m_{\pi}/M\right)^2 }$, where $M$ is the  mass of the ground state, $m_\pi$ is the pion mass,  $\Delta M^S=M\left(\Xi_{cc}(1S1s(3/2)\right) - M\left(\Xi_{cc}(1S1s(1/2)\right)=M\left(\Xi_{cc}(1S1s(3/2)\right) - M$, and $\Delta M^{PS}$  is the mass difference between $1P1s(3/2)$ and $1S1s(3/2)$ states: $\Delta M^{PS}=M\left(\Xi_{cc}(1P1s(3/2)\right) - M\left(\Xi_{cc}(1S1s(3/2)\right) $.}  Here is it worth to note, that the $1S1s$-multiplet transition  can occur via a photon emission only, because  the value of   mass splitting in $1S1s$-doublet is about $100\div 130$~MeV~\cite{Ebert:2002ig,Brambilla:2005yk, Fleming:2005pd, Kiselev:2017eic}, i.e. it is less than the pion mass.   
 
In $\Omega_{cc}$ spectrum single-pion transitions break the isospin symmetry and therefore, if kinematically possible, the $\Omega_{cc}$ excitations decay into the $\Xi_{cc}$ ground state via kaon emission.  A special case is the first P-wave diquark excitation of $\Omega_{cc}$. The single-pion transitions are strongly suppressed due to  isospin symmetry breaking, and the single-kaon transitions are kinematically forbidden.  As a result, for such states the hadronic mode does not dominate towards the electromagnetic one (see \cite{Ma:2017nik} and \cite{Dai:2000hza}).

In the conclusion of this chapter, it should be noted that the decays of excited doubly charmed baryons are rather poorly studied, and therefore more detailed studies are required.

\section{Conclusions}

In this study we calculate the relative yields of $S$ and $P$ wave $\rho$-excitations of doubly charmed baryons at LHC. 
The observation of  narrow metastable $P$ wave $\rho$-excitations of $\Xi_{cc}$ in the decay mode $\Xi_{cc}\pi$ is rather challenging because of small yield of such states, which is about 3\%. On the contrary,  the structure corresponded to  the decays $\Xi_{cc}\left(2S,3S\right) \to \Xi_{cc}\pi$   should be definitely observed in Run III.  Indeed, about 50 \%  of $\Xi_{cc}^{++}$ baryons come from  $S$-wave excitations. Therefore from $\sim 300$ $\Xi_{cc}^{++}$ observed at LHCb  $\sim 100$ baryons are the products of decays with charged pion $\Xi_{cc}^{+}\left(2S,3S\right) \to \Xi_{cc}^{++}\pi^-$ and  $\sim 10$ baryons come from the decays with charged kaon $\Omega_{cc}^+ \left(2S, 3S\right) \to \Xi_{cc}^{++} K^-$ . This is why we think, that Run III with large luminosity will provide a great opportunity to observe excited $S$-wave states of  $\Xi_{cc}^{+}$ and  $\Omega_{cc}^+$ baryons.

Authors thank V. Galkin for help and fruitful discussion.
A. Berezhnoy and I. Belov  acknowledge the support from  "Basis" Foundation (grants 17-12-244-1 and 17-12-244-41).

\bibliography{dhb-litr} 

\end{document}